\documentclass[doublecol]{epl2}
\usepackage{amssymb}
\usepackage{amsmath}
\usepackage{graphicx}
\usepackage{bm}
\makeatletter

\begin{document}

\title{Quasiparticles in the Pseudogap Phase of Underdoped Cuprate}
\author{Kai-Yu Yang\inst{1,3}, H.-B. Yang\inst{2}, P. D. Johnson\inst{2}, T. M. Rice\inst{1,3}, Fu-Chun Zhang\inst{3}}
\institute{
\inst{1} Institut f$\ddot{u}$r Theoretische Physik, ETH Z$\ddot{u}$rich,CH-8093 Z$\ddot{u}$rich, Switzerland \\
\inst{2} Condensed Matter Physics and Materials Science Department, Brookhaven National Laboratory, Upton, NY 11973, USA\\
\inst{3} Center for Theoretical and Computational Physics and
Department of Physics, The University of Hong Kong, Hong Kong SAR,
China }


\abstract{
Recent angle resolved photoemission \cite{yang-nature-08} and
scanning tunneling microscopy \cite{kohsaka-nature-08} measurements on underdoped cuprates have yielded
new spectroscopic information on quasiparticles
in the pseudogap phase. New features of the normal state such as
particle-hole asymmetry, maxima in the energy dispersion and accompanying drops in the spectral weight of quasiparticles agree with the ansatz of Yang
\textit{et al.} for the single particle propagator in the
pseudogap phase. The coherent quasiparticle dispersion and reduced asymmetry in the
tunneling density of states in the superconducting state can also be described by this propagator.
}

\date{\today}
\maketitle

  The anomalous electronic properties of the pseudogap phase in underdoped cuprates continue to challenge theory. Angle resolved
  photoemission (ARPES) measurements of the Fermi surface in the pseudogap state show only disconnected Fermi arcs centered on the nodal directions.
  \cite{arpes-rmp-03, kanigel-naturephys-06, norman-prb-07} A recent high resolution ARPES study on underdoped BSCCO by Yang and collaborators \cite{yang-nature-08} examined the spectra below
  and above the chemical potential $\mu$ by carefully dividing out the Fermi function.
They found that the spectra displayed particle-hole asymmetry away
from the nodal direction in the pseudogap state but not in optimal
doped samples, so that the asymmetry is a property of the
pseudogap phase. Scanning tunneling microscopy (STM) by Kohsaka
and collaborators \cite{kohsaka-nature-08} has determined the
coherent quasiparticle dispersion in a wide range of hole
densities in the superconducting state, by an ingenious analysis
of the voltage dependent spatial interference patterns.
\cite{wang-prb-03} The asymmetry in the tunneling density of
states measured by STM is weaker due to the inherent particle-hole
symmetry of superconductivity. Recent angle integrated
photoemission (AIPES) experiments on underdoped samples
\cite{hashimoto-arxiv-08}found the density of states varied
linearly with energy in contrast to the constant value that
appears at a band edge in two dimensions. In this letter we
analyze ARPES, STM and AIPES results using an ansatz for the
single particle
propagator proposed earlier by three of us. 
Key properties of the quasiparticles (QP) in this propagator, such
as maxima in the energy dispersion in the normal pseudogap state,
the connection between the QP dispersion in the superconducting
state and the hole density, are directly confirmed by these
experiments. An alternative explanation of particle-hole asymmetry
in ARPES experiments has been proposed by Anderson
\cite{Anderson-arXiv-08}.

  As the hole density, $x$, crosses the underdoped pseudogap region of the phase diagram, the cuprates crossover from a full Fermi surface metal at overdoping to a Mott
  insulator at stoichiometry. A Mott insulating state is driven by strong onsite Coulomb repulsion and not \textit{per se} by translational symmetry breaking. The
  stoichiometric underdoped cuprate YBa$_2$Cu$_4$O$_8$, shows no sign of a charge or spin density modulation on either Cu or O sites. \cite{tomeno-prb-94} It is therefore
  desirable to examine theories where the truncated Fermi surface appears as a precursor to the Mott state without translational symmetry breaking. An example is the 2D array
  of 2-leg Hubbard ladders studied by Konik, Rice and Tsvelik (KRT). \cite{KRT-prl-06} They showed that lines of zeros in the propagator $G^{R}(\boldsymbol{k},\omega=0,x)$ associated with
  the charge gap at half-filling, enclosing a commensurate area of 1 el/site, do not move with light doping. The Luttinger Sum Rule on the area enclosed by the condition
  $Re[G^{R}(\boldsymbol{k},\omega=0,x)]>0$ is fulfilled with the Fermi surface truncated to small hole pockets.

A relevant question is whether similar behavior can occur in a fully 2D system. Honerkamp and coworkers \cite{honerkamp-prb-01} argued for such behavior from the similarities in the flow of the response
  functions in functional renormalization group analyses of Hubbard models near half-filling.
  Yang, Rice and Zhang \cite{yang-prb-06} (YRZ) proposed an adaptation of the KRT approach to the lightly doped $t$-$J$ model. They introduced a self-energy, $ \Sigma^{R}(\boldsymbol{k},
  \omega, x)$, which diverges at $\omega = 0$ on a surface spanned by elastic particle-particle umklapp scattering analogous to the behavior of the ladder model. In the
  2D square lattice this umklapp surface is a diamond connecting antinodal points on the Brillouin zone boundary. Note this umklapp surface appears as the energy gap
  surface also in the case of wider Hubbard ladders with more than 2 legs. \cite{lehur-review}

YRZ took over the form for the gap function $\Delta^{R}(\boldsymbol{k}, x)$ from the renormalized mean field
  theory of Zhang \textit{et al.} \cite{RMFT} for the resonant valence bond (RVB) state \cite{Anderson-JPCM-04, Ogata-RPP-08} of the strong coupling 2D $t$-$J$ model and proposed an ansatz for the single particle
  propagator:
\begin{eqnarray}
G^{R}(\boldsymbol{k}, \omega, x)={g^{t}(x)}/[{\omega -\xi_{\boldsymbol{k}}-\Sigma^{R}(\boldsymbol{k}, \omega, x) }]+G_{inc}
\label{eq:PG1}
\end{eqnarray}
where 
$\Sigma^{R}(\boldsymbol{k},\omega,x)=\left\vert
\Delta^{R}(\boldsymbol{k} ,x)\right\vert ^{2}/(\omega
+\xi^{(0)}_{\boldsymbol{k}})$ is the RVB self-energy. The energy
$\xi^{(0)}_{\boldsymbol{k}}=-2t(x)(\cos
k_{x}+\cos k_{y})$ vanishes on the umklapp surface.
The renormalized dispersion $\xi_{\boldsymbol{k}}
=\xi^{0}_{\boldsymbol{k}}-4t^{\prime }(x)\cos k_{x}\cos k_{y}
-2t^{\prime \prime }(x)(\cos 2k_{x}+\cos 2k_{y})-\mu$
 includes hoppings out to 3$^{\text{rd}}$ nearest neighbor with hole density
dependent coefficients $t(x)=g^{t}(x)t_{0}+3g^{s}(x)J\chi
/8$, $t^{\prime }(x)=g^{t}(x)t_{0}^{\prime }$, and $t^{\prime
\prime }(x)=g^{t}(x)t_{0}^{\prime \prime }$. The Gutzwiller factors $g^{t}(x)=2x/\left( 1+x\right)$ and
$g^{s}(x)=4/(1+x)^{2}$ account for the no-double occupation projection on the kinetic
and superexchange terms in the $t$-$J$ model. The RVB energy gap takes the form $\Delta^{R}(\boldsymbol{k},x)
=\Delta^{(0)}(x)(\cos k_{x}-\cos k_{y})$. $\chi$ corresponds
to the weakly doping dependent homogeneous amplitude
$< c_{i,\sigma }^{\dagger
}c_{j,\sigma }> $. Parameters are quoted in Table \ref{tab:parameter}.

\begin{table} [t]
\caption{ The bare hopping parameters $t^{\prime}_{0}$, $t^{\prime\prime}_{0}$ and the superexchange $J$ are quoted in units of the nn value $t_{0}=0.3eV$. The hole density
  determines the pocket area and is consistent with the value determined
  independently\cite{yang-nature-08}. \newline}
\begin{tabular} {|c|c|c|c|c|c|c|}  \hline\hline
$t^{\prime}_{0}$ & $t^{\prime\prime}_{0}$ & $J$ & $x$ &$\Delta^{0}$& $\Delta^{0}(x)$&$\chi$ \\ 
-0.3       & 0.2 & 1/3 &0.12 &0.5& $\Delta^{0}(1-\frac{x}{0.2})$&0.338 \\
\hline \hline
\end{tabular}
\label{tab:parameter}
\end{table}

  As discussed by YRZ the Fermi surface in the pseudogap regime with divergent $G^{R}(\boldsymbol{k}, \omega=0, x)$ has 4 pockets centered on the nodal directions
  enclosing a total area related to the hole density. \cite{yang-prb-06} ARPES experiments  \cite{arpes-rmp-03, kanigel-naturephys-06, norman-prb-07, yang-nature-08} show arcs
rather than closed pockets, but as YRZ pointed out the
QP weight in $G^{R}$ varies rapidly with very small
values on the back sides of the pockets, which can account for the
failure to observe these parts of the pockets. Many experimental features,
\textit{e.g.} the slow variation of the nodal Fermi velocity
$v_{F}$ and Drude weight scaling with hole density, are reproduced
by the YRZ propagator. \cite{yang-prb-06} The umklapp surface along which the energy
gap opens up, lies above $\mu$, causing
particle-hole asymmetry in the QP spectra in
  the pseudogap state. The spectra are symmetric only along the nodal directions where $\Delta^{R}(\boldsymbol{k},x)=0$, and the asymmetry increases away from the nodal directions.

\begin{figure}[t]
\centerline{\includegraphics[width = 7.5cm, height = 10.0cm, angle
= 0]{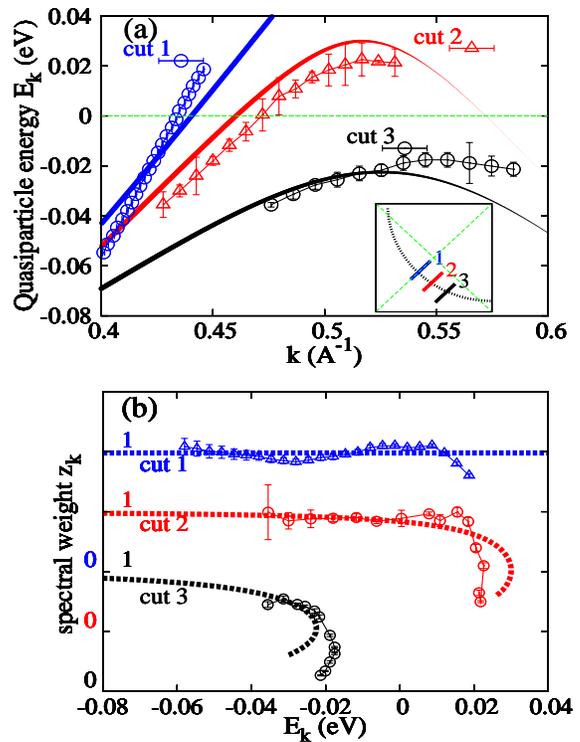}} \caption[]{
(Color online) Comparisons between (a) QP
dispersion $E_{\boldsymbol{k}}$, (b) spectral weight
$z_{\boldsymbol{k}}$, from the YRZ propagator and the values
obtained from the ARPES results by Yang \textit{et al.}
\cite{yang-nature-08} Error bars reflect the uncertainties in the
fitting procedures; $z_{\boldsymbol{k}}$ at low energies in cuts
2, 3 have large error bars due to uncertainties in the choice of
the rising background.} \label{fig:parameter}
\end{figure}

\begin{figure}[t]
\centerline{\includegraphics[width = 10.5cm, height = 9.5cm,
angle=270]{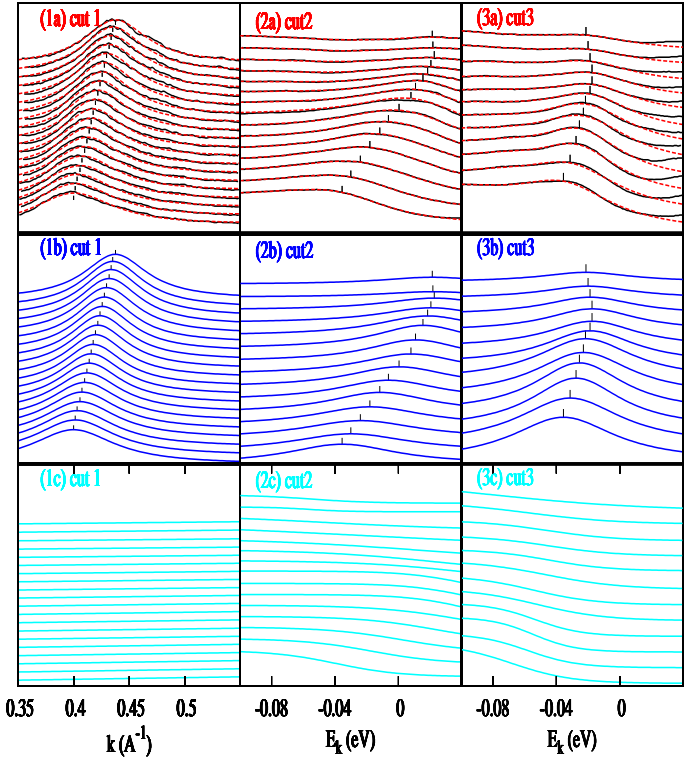}} \caption[]{ (Color online) Fits to the ARPES
spectra along the cuts (1-3) (see inset of
Fig.\ref{fig:parameter}) \cite{yang-nature-08}. In (1a) the MDC of
cuts 1 and in (2/3a) the EDC of cuts 2 and 3 are shown, the
experimental data are the solid black lines, and the fitted
$A(\boldsymbol{k},\omega)$ are the dashed red lines. Individual
components of the fits are displayed in (b) for QP Lorentzian
peaks and (c) for smooth background
$A^{B}(\boldsymbol{k},\omega)$.
} \label{fig:fit}
\end{figure}

  The spectra observed by Yang \textit{et al} \cite{yang-nature-08} along three cuts (see inset in Fig.\ref{fig:parameter}) in the pseudogap phase ($T$=$140$K) are shown in the Brillouin zone in Fig.\ref{fig:fit}(a). They are fits to a broadened Lorentzian plus a background term to be defined below. The fitting uses an
  implementation of the nonlinear least-square Marquardt-Levenberg algorithm. We determine the QP dispersion, $E_{\boldsymbol{k}}$, weight $z_{\boldsymbol{k}}$, and inverse lifetime $\gamma_{\boldsymbol{k}}$ for each $\boldsymbol{k}$-value from the form.
\begin{eqnarray}
A(\boldsymbol{k}, \omega) &=& Im [{z_{\boldsymbol{k}}}/ ({\omega -E_{\boldsymbol{k}} + i \gamma_{\boldsymbol{k}} })] + A^{B}(\boldsymbol{k}, \omega) \label{EQ:fitting}
\end{eqnarray}
For the background, momentum distribution curves (MDC) of cut 1 are well fitted with an almost
constant background $A^{B}(\boldsymbol{k},\omega)$, for energy distribution curves (EDC) of cuts 2 and 3 we use a broadened step function which rises at lower energies well below $\mu$,
\begin{eqnarray}
A^{B}(\boldsymbol{k}, \omega) &=& y_{\boldsymbol{k}} \left
(\tanh{[\lambda_{\boldsymbol{k}} (\Omega_{\boldsymbol{k}}-\omega)}]+1
\right ) + \alpha_{\boldsymbol{k}} \label{EQ:bg}
\end{eqnarray}
 The fitted MDC(EDCs) for the three cuts are
shown in the upper panels of Fig.\ref{fig:fit}(a), and the two
components, the Lorentzian peak and background,
in the lower panels (b/c). For cut 1 along the nodal directions we note
that the fitting to a broadened Lorentzian with an almost constant background works well for all energies.
For cuts 2-3 EDC fitting, as $\boldsymbol{k}$ moves away
from $\boldsymbol{k}_{F}$, the background component
$A^{B}(\boldsymbol{k},\omega)$ increases at lower energies. We
believe this increase arises from contamination from adjacent $\boldsymbol{k}$-values, due to the
enhanced broadening at low energies. Since our interest focusses
on energies near $\mu$ we believe this choice
of background which is constant near $mu$ is reasonable.

 Fig.\ref{fig:parameter} shows the results for the QP properties of interest. The QP dispersion,
$E_{\boldsymbol{k}}$ in Fig.\ref{fig:parameter}(a) rises linearly
with $\boldsymbol{k}$ along the nodal cut 1. Particle-hole asymmetry is
evident in cuts 2 (3) with a maxima in $E_{\boldsymbol{k}}$
lying above (below) $\mu$, respectively. Note that these maxima are not at the boundary of the reduced Brillouin zone as the case would be in the presence of a broken
translation symmetry, a point emphasized by Yang \textit{et al.} \cite{yang-nature-08}
 The QP weight, $z_{\boldsymbol{k}}$, represented in the dispersion plot as the line width, is plotted in
Fig.\ref{fig:parameter}(b). We see an almost constant
$z_{\boldsymbol{k}}$ for cut 1 along the nodal direction, but a
fall off at the maximum in $E_{\boldsymbol{k}}$ for cuts 2 and 3
in agreement with the YRZ ansatz. The inverse lifetime
$\gamma_{\boldsymbol{k}}$ is essentially constant indicating
substantial inelastic scattering at this elevated temperature of
140K. To summarize with reasonable parameters the
phenomenological YRZ propagator gives a good fit to the
asymmetry of ARPES spectra and describes both the maxima in the
QP dispersion, $E_{\boldsymbol{k}}$, away from the
nodal direction and the accompanying drop off in the weight,
$z_{\boldsymbol{k}}$.

\begin{figure}[bt]
\centerline{\includegraphics[width=8.0cm, height = 8.0cm, angle
=270]
{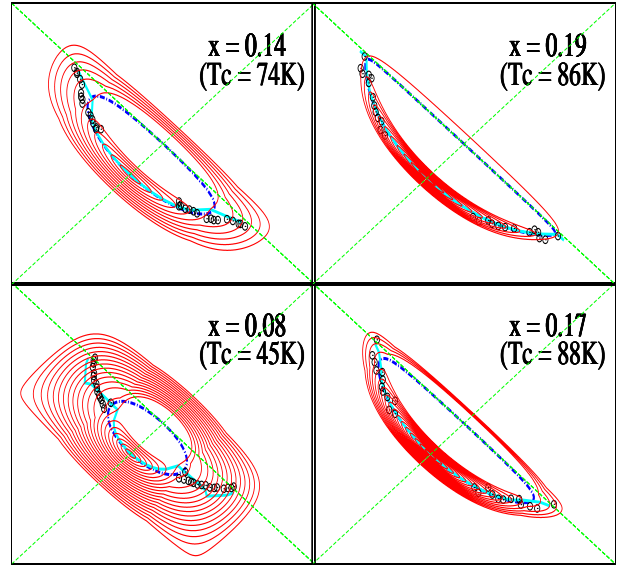}
}

\caption[]{(Color online) The red curves are the contours of the
constant QP energy below
 and close to the Fermi energy. The black dots are the turning points deduced from STM interference data. The dashed blue lines are the hole
 Fermi pocket in the normal pseudogap state. The cyan curves are the turning points of the iso-energy contour of the corresponding band in the SC state. Note only the
 turning point of the iso-energy contour with the largest spectral weight are shown.} \label{fig:FS}
\end{figure}

\begin{figure}[bt]
\centerline{\includegraphics[width=8.0cm, height = 8.0cm, angle=0]{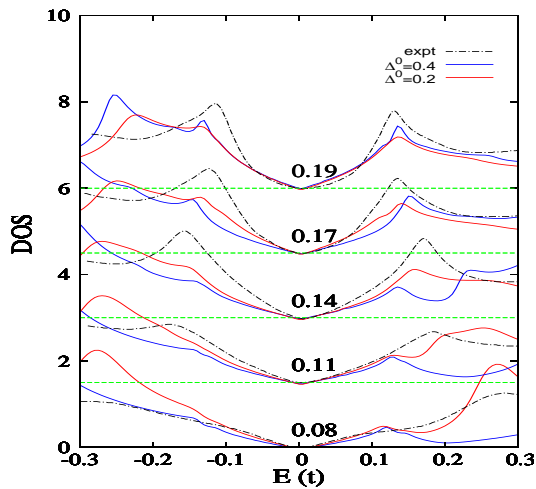}}
\caption[]{(Color online) The density of states in the
superconducting state for various dopings. The d-wave SC gap is
chosen to saturate at ends of Fermi pockets at a fixed value
$\Delta_{S} |_{\text{max}}=0.14$ to agree with experiments. Two choices
for the RVB gap $\Delta_{0}$ are displayed (red $\Delta_{0}=0.2$,
blue $\Delta_{0}=0.5$). The dashed black curves show the STM data
\cite{kohsaka-nature-08} in units of $t_{0}$ (300meV). The inverse of
lifetime $\Gamma = \alpha |E_{QP}|$ with $\alpha = 0.1$ is used.} \label{fig:DOS}
\end{figure}

\begin{figure}[bt]
\centerline{\includegraphics[width=8.0cm, height = 8.0cm, angle=0]{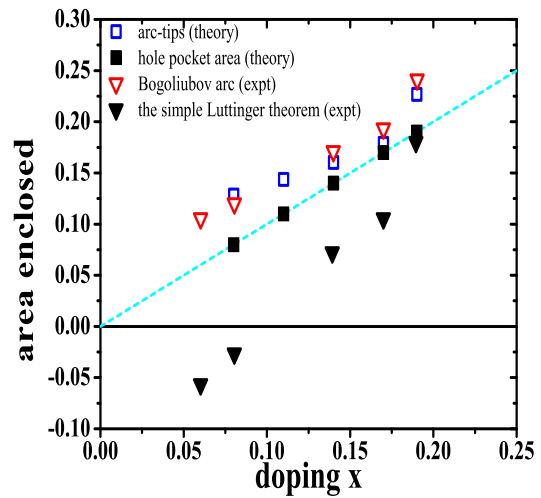}}
\caption[]{(Color online): Hole densities implied by areas enclosed by different
interpretations of the QP dispersion shown in
Fig.\ref{fig:FS}, and dispersion extrapolated using LDA band
structure ($\blacktriangledown$from Ref\cite{kohsaka-nature-08}),
hole pockets determined by turning points in Fig.\ref{fig:FS} and
lines connecting antinodal points ($\bigtriangledown$ from
Ref\cite{kohsaka-nature-08}, $\Box$ from this work), cyan line and
$\blacksquare$ from the YRZ propagator for the Fermi pocket. .} \label{fig:area_enclosed}
\end{figure}

In their original discussion of the ARPES spectra, Yang \textit{et
al.} \cite{yang-nature-08} presented results for a fourth scan
closer to the antinodal point (cut 2 in Fig.4 of Ref.
\cite{yang-nature-08}) than scans (1-3) in Fig. 3 of Ref.
\cite{yang-nature-08}. This showed particle-hole symmetry in the
accessible energy window near the Fermi energy ($\lesssim 40$
meV). Note, this energy window is too small to see the actual gap
on the particle side. However, the observation that the minimum in the gap is located right at the chemical potential led Yang \textit{et al.}
\cite{yang-nature-08} to suggest that the pseudogap at the
antinodal regions has a pairing origin, in agreement with the
proposals of Kanigel \textit{et al.} \cite{kanigel-naturephys-06} and
Anderson \cite{Anderson-arXiv-08}. In a standard pairing theory
the energy gap opens up along the Fermi surface whereas in the YRZ
propagator the gap opens up at a fixed surface in
$\boldsymbol{k}$- space. As such some of use (YRZ) argue that this occurs due to the
presence of umklapp particle-particle scattering in addition to
the Cooper channel, similar to the case of the half-filled Hubbard
2-leg ladder. \cite{honerkamp-prb-01} As a consequence of this
difference, the RVB energy gap, $\Delta^{R}$, acts as a charge gap
and only the quasiparticle excitations at the Fermi pockets
contribute to the reduced Drude weight and reduced coherent
stiffness in the superconducting state. That are characteristic of
the pseudogap state. \cite{Basov-RMP-05}

At this point we wish to comment the apparent contradiction between the hole Fermi pockets associated with the YRZ form and the electron Fermi pockets deduced from recent quantum oscillation experiments on underdoped YBCO samples. \cite{leyrand-nature-07}. These experiments are performed at high magnetic fields and low temperatures whereas the ARPES experiments of Yang \textit{et al.} \cite{yang-nature-08} were made at zero field and at a relatively high temperature of 140K. Measurements of the Hall constant on the YBCO samples show a hole-like behavior under the latter conditions but electron-like Hall constant under the conditions of the quantum oscillation experiments \cite{taillefer-nature-07}. This suggests that the sign change in the Hall constant and the electron pockets, arise from a reconstruction of the Fermi surface \cite{millis-prb-07} at high magnetic fields, possibly related to the similar sign change in the Hall constant that accompanies the superlattice charge and spin order in the static stripe phase.  \cite{taillefer-nature-07}

The recent STM results of Kohsaka \textit{et al.}, \cite{kohsaka-nature-08} requires a
generalization of the YRZ propagator to the superconducting state.
 Originally YRZ added a self-energy
term in $G^{R}(\boldsymbol{k},\omega, x)$. \cite{yang-prb-06}
However, during this work we realized that this is appropriate
only in the presence of particle-hole symmetry. It is more
convenient to rewrite Eq.\ref{eq:PG1}
\begin{eqnarray}
G^{R}(\boldsymbol{k},\omega, x) =
\sum_{\alpha=\pm} {W^{\alpha}_{\boldsymbol{k}}}/({\omega-E^{\alpha}_{\boldsymbol{k}}}) +G_{inc}
\label{eq:PG}
\end{eqnarray}
where $E_{\boldsymbol{k}}^{\pm }= \bar{\xi}_{\boldsymbol{k}}\pm
\sqrt{\bar{\xi}_{\boldsymbol{k}}^{2}+\epsilon
_{\boldsymbol{k}}^{2}}$, $\bar{\xi}_{\boldsymbol{k}}=(\xi _{\boldsymbol{k}}-\xi
_{\boldsymbol{k}}^{\left( 0\right) })/2$, $\epsilon _{\boldsymbol{k}}^{2}=\xi
_{\boldsymbol{k}}\xi _{\boldsymbol{k}}^{\left( 0\right)
}+\left\vert \Delta _{\boldsymbol{k}}^{R}\right\vert ^{2}$ and $(W^{\pm}_{{\boldsymbol{k}}})^{-1}=1+ {\left\vert \Delta
_{\boldsymbol{k}}^{R}\right\vert ^{2}}/{(
E_{\boldsymbol{k}}^{\pm }+\xi_{\boldsymbol{k}}^{\left( 0\right)
}) ^{2}}$. The generalization to
the superconducting state is straightforward by treating the system
as a two-band d-wave superconductor with gap
$\Delta_{S}(\boldsymbol{k}, x) = \Delta_{S}(x)
[\cos(k_{x})-\cos(k_{y})]$ (for simplicity we use a single SC gap)
leading to a propagator
\begin{eqnarray}
G^{S}(\boldsymbol{k},\omega,x)=
\sum_{\alpha=\pm}  {W^{\alpha}_{\boldsymbol{k}}}/[
{\omega-E^{\alpha}_{\boldsymbol{k}}-|\Delta_{S}(\boldsymbol{k},
x)|^{2}/(\omega+E^{\alpha}_{\boldsymbol{k}})}]
\label{Eq:SC}
\end{eqnarray}
with four QPs
$E_{i,\boldsymbol{k}}=\pm\sqrt{(E^{\alpha}_{\boldsymbol{k}})^{2}+\Delta_{S}(\boldsymbol{k},
x)^{2}}$ with spectral weight $z_{i,\boldsymbol{k}}$.

The superconducting gap opens up along the hole pockets. Kohsaka
\textit{et al} \cite{kohsaka-nature-08} derived the coherent QP dispersion in
the superconducting state from the evolution of the
interference pattern in the STM spectra measured in a wide surface
area as the tunneling voltage is changed in BSCCO samples,
ranging from near optimal doping ($x=0.19$) to strongly underdoped
($x=0.06$). This determines a set of $\boldsymbol{q}$-vectors connecting
the turning points in the iso-energy contours of the
QP dispersion. \cite{wang-prb-03, bascones-prb-08} These turning points can be compared to the values obtained from Eq.\ref{Eq:SC}. We use the same values of the parameters $t_{0}$,
$t^{\prime}_{0}$, $t^{\prime\prime}_{0}$ and $J$ as fits to the ARPES spectra at $x=0.12$.
Following Zhang \textit{et al} \cite{RMFT} we assume the RVB gap value
$\Delta^{0}(x)$$[=\Delta^{0}(1-x/x_{c})]$ drops linearly with $x$ with a critical doping $x_{c}=0.20$. Note the value $\Delta^{0}=0.5$ is used
in fitting to ARPES results on a sample with $x=0.12$. In fitting to the STM results we shall allow for the possibility that the superconducting $\Delta_{S}$ and RVB
$\Delta^{0}$ gaps are not independent and that $\Delta^{0}$ may change in the superconducting state from its value in the normal state.

The remaining issue is the form for the superconducting gaps
$\Delta_{S}(\boldsymbol{k}, x)$. Generally we expect that
$\Delta_{S}(\boldsymbol{k}, x)$ will not continue to grow as
$\boldsymbol{k}$ moves beyond the Fermi pockets and the normal
state QPs drop below the Fermi energy. Therefore we assume a
d-wave form for $\Delta_{S}(\boldsymbol{k}, x) = \Delta^{0}(x) (\cos k_{x} -\cos k_{y})$ only for
$\boldsymbol{k}$-values on the pockets and set
$\Delta_{S}(\boldsymbol{k}, x)$ to be constant beyond the pockets.
This results in a two-gap description \cite{deutcher,
le-tacon-naturephys-06, valenzuela-prl-07, boyer-nature-phys-07, lee-nature-07, hufner-rpp-08, Nicol-08, blanc-arxiv-09, kondo-nature-09} with the
antinodal energy gap dominated by the RVB gap at underdoping while
the superconducting gap along the hole pockets is more important
near optimal doping. We display in Fig.\ref{fig:FS} the evolution
of the contours of constant QP energy near $\mu$. Note the strong
variation in QP weight discussed earlier, means that only the
outer contours closest to the zone center are relevant. The
turning points of these contours are determined by the maxima in
the inverse velocity $\left( dE(\boldsymbol{k})/dk_{\bot}
\right)^{-1}$ along these contours. The comparison to the STM
results is
illustrated in Fig.\ref{fig:FS} for 4 representative hole densities
. We see that the overall agreement is quite satisfactory.
Note that the contours defined by the turning
points are slightly larger than the underlying hole Fermi pockets
and enclose a slightly larger area. This causes the hole density implied by the turning point contour to be slightly larger than
the actual hole density. This discrepancy is evident in Fig.3(a)
of Kohsaka \textit{et al.} \cite{kohsaka-nature-08}. The relevant areas are plotted
in Fig.\ref{fig:area_enclosed}.

Note that in the above analysis we used a standard d-wave form for
the superconducting gap, $\Delta_{S}(\boldsymbol{k},x) =
\Delta^{0}_{S}(x) (\cos k_{x} - \cos k_{y})$ as observed in ARPES.
Kohsaka \textit{et al.} in their STM experiments found a different
angular dependence with a strongly reduced gap near to the nodal
directions. The origin of this discrepancy is not clear at
present. One possibility could be enhanced forward impurity
scattering in the samples used in the STM experiments which Haas
\textit{et al.} showed can lead to a gap suppression near to the
nodal directions. \cite{haas-prb-97} However since both techniques
are sensitive only to the surface layers it is far from clear that
such an explanation is viable.

Kohsaka \textit{et al.} \cite{kohsaka-nature-08} report an abrupt
end to the coherent QP dispersion in the STM experiments when
$\boldsymbol{k}$ reaches the umklapp surface. The YRZ propagator
contains coherent QP also in the gapped antinodal regions. QP in
these regions, however, should be sensitive to the strong local
variations in the hole density and the accompanying large
variations in the RVB energy gap, $\Delta^{0}_{R}(x)$.
\cite{wise-naturephy-09} This may well lead to localization of the
corresponding  QP states which lie at energies away from the
chemical potential $\mu$.  The constant value of $\mu$ makes the
Bogoliubov QP states associated with the superconducting gap much
less sensitive to disorder. This difference offers a plausible
explanation for the change from propagating to localized states
that Kohsaka \textit{et al.} \cite{kohsaka-nature-08} propose
takes place as $\boldsymbol{k}$ reaches the umklapp surface.


Kohsaka \textit{et al.} \cite{kohsaka-nature-08} reported
characteristic tunneling spectra, which we can compare to the DOS
obtained from the superconducting propagator in Eq.\ref{Eq:SC}
$N(\omega) = \sum_{i,\boldsymbol{k}} z_{i,\boldsymbol{k}}
\delta_{E_{i,\boldsymbol{k}}-\omega}$, as shown in
Fig.\ref{fig:DOS}. The experimental results are shown in
Fig.\ref{fig:DOS} as the dashed black curves. At higher hole
densities, $0.1<x<0.19$, the spectra are dominated by the maximum
superconducting gap which is larger than the RVB gap,
$\Delta^{0}(x)$. As a result the QP bands near the Fermi energy
are all split by $\Delta_{S}(\boldsymbol{k})$ with symmetric low
energy DOS. At lower hole density the RVB gap $\Delta^{0}(x)$
rises and exceeds the maximum of $\Delta_{S}(\boldsymbol{k})$
leading to two structures in the DOS. One at lower energy related
to $\Delta_{S}(\boldsymbol{k},x)$ at the tips of arc, and  second
at higher energies associated with the RVB gap.
If we keep the RVB gap $\Delta^{0}$ to the value used in the ARPES fits, the DOS shown in blue in Fig.\ref{fig:DOS} displays a much stronger particle-hole
asymmetry than the experiments. However, if we reduce the value of $\Delta^{0}$ to 0.2, the agreement is much improved, suggesting that the two gaps
are not independent of each other. Note that the total gap near the antinodal points does not change so much.
 We do not claim a quantitative fit to the STM DOS but the main features are
reproduced at least qualitatively, by the YRZ ansatz.

\begin{figure}[bt]
\centerline{\includegraphics[width=7.0cm, height = 10.0cm,
angle=270]{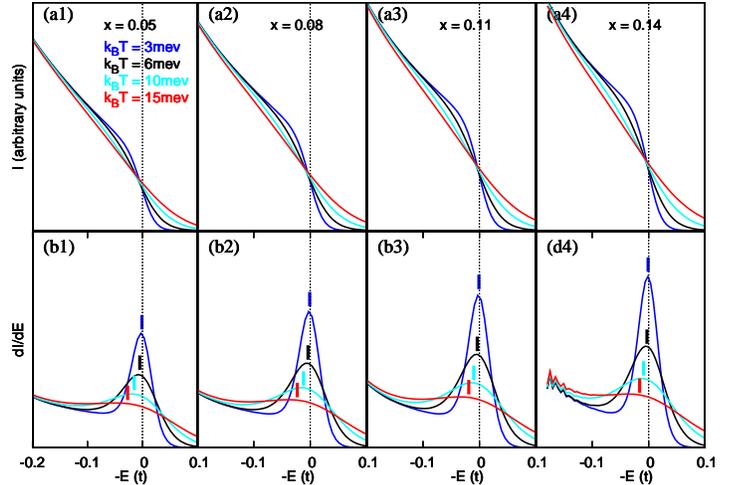}} \caption[]{(Color online): The upper
panels are the angle integrated spectra $I(E, T, x)$ near the
chemical potential (at zero temperature) for various doping (0.05
- 0.14).  The lower panels are the first order derivative of $I(E,
T,x)$, with the peak position $E_{p}$ shifting towards negative
electron energy (\textit{i.e.} positive hole energy), opposite to
the shift in the chemical potential. Blue, black, cyan and red
curves are for the temperatures $k_{B}T = 3, 6, 10, 15 $ meV. The
same parameters as shown in Table 1 are used, except $\Delta^{0} =
0.4$. Experimental data for La$_{2}$CuO$_{4}$ can be found in
Fig.xxx of Hashimoto \textit{et al} \cite{hashimoto-arxiv-08}.}
\label{fig:dos_PG_T}
\end{figure}

\begin{figure}[bt]
\centerline{\includegraphics[width=5.5cm, height = 7.0cm, angle=270]{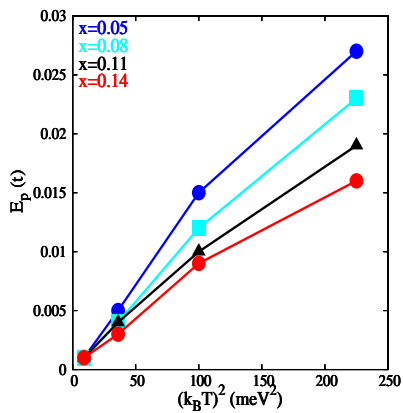}}
\caption[]{(Color online): The shifting value of the peak position $E_{p}$ of the derivative of $dI/dE$ towards positive hole energy for varous doping and temperature. It shows a clear quadratic temperature dependence.} \label{fig:peak}
\end{figure}

 Next we turn to recent angle integrated photoemission (AIPES) experiments on hole doped La$_{2}$CuO$_{4}$
 which showed several anomalous features at underdoping in the normal pseudogap phases.
 First the total DOS (\textit{i.e.} angle integrated) show an approximate linear dependence on the hole energy, $E$,
 with only a small quadratic correction which grew with increasing $x$. This behavior contrasts with the constant DOS
 expected from a small hole pocket in a valence band in two dimensions. Secondly, the shift of the peak in the first
 derivative of the angle integrated spectra ($I(E, T, x)$), $dI/dE \propto T^{2}/T_{coh}$ approximately and moves to positive
 hole energies (not negative). The value of $T_{coh} (x)$ increases with increasing, $x$. Interestingly,
 these anomalous features are reproduced by the YRZ model dispersion for quasiparticles. In Fig. \ref{fig:dos_PG_T} (a)
 we show the results for the angle integrated spectra $I(E,T,x)$ calculated by multiplying the total DOS for quasiparticles by the
 Fermi function. An approximate linear dependence on $E$ is evident. In Fig. \ref{fig:dos_PG_T} (b) we show that the peak $E_{p}$ of
 the derivative $\mid dI /dE \mid$ moves to positive hole energies opposite to the negative shift of the chemical potential,
 with a roughly quadratic dependence on the temperature, Fig. \ref{fig:peak}. The resulting characteristic coherence temperature $T_{coh}(x)$
 increases with increasing $x$. The key anomalous properties of the
AIPES spectra are well reproduced by the YRZ form.

The YRZ ansatz for the single particle propagator explains the
anomalous properties of the underdoped pseudogap phase as a
precursor to the Mott insulating state at stoichiometry. The
recent spectroscopic measurements on underdoped BSCCO using ARPES
and STM have shed new light on the evolution of the QP properties.
In this paper we have shown that the YRZ ansatz provides the basis
to understand the key features of the new experimental data such
as particle-hole asymmetry, energy dispersion and wave vector
dependent spectral weight of normal state QP as well as the
coherent QP dispersion and DOS in the superconducting state over a
range of hole densities.

We thank Wei-Qiang Chen, Seamus Davis,  Carsten Honerkamp, Manfred Sigrist and Alexei Tsvelik for
discussions. Support from the MANEP program of the Swiss National
funds (K.-Y. Y. and T. M. R.), the US Department of Energy under Contract No. DE-AC02-98CH10886 (H.-B. Y.
and P. D. J.) and RGC grant of HKSAR (K.-Y. Y and F. C. Z.) is
gratefully acknowledged.

\end{document}